\documentclass[10pt,oneside]{amsart}
\usepackage{amstext,amsmath,amsfonts,amscd,amsthm}
\usepackage{cite}
\input epsf

\theoremstyle{plain}
\newtheorem{theorem}{Theorem}

\newtheorem{proposition}[theorem]{Proposition}

\theoremstyle{definition}
\newtheorem{definition}{Definition}
\newtheorem{conjecture}{Conjecture}
\newtheorem{observation}{Observation}
\newtheorem{example}{Example}

\theoremstyle{remark}

\newtheorem{note}{Note}
\begin{document}

\title{Spectral Geometry and Causality}

\author[T.Kopf]{Tom\'{a}\v{s} Kopf\\
Department of Physics, University of Alberta\\
Edmonton, Alberta, Canada, T6G 2J1}

\begin{abstract}
For a physical interpretation of a theory of
quantum gravity, it is necessary to recover
classical spacetime, at least approximately.
However, quantum gravity may eventually provide
classical spacetimes by giving spectral data
similar to those appearing in noncommutative
geometry, rather than by giving directly a
spacetime manifold. It is shown 
that a globally hyperbolic Lorentzian manifold
can be given by spectral data.  A new phenomenon
in the context of spectral geometry is observed:
causal relationships. The employment of the
causal relationships of spectral data is shown to
lead to a highly efficient description of
Lorentzian manifolds, indicating the possible
usefulness of this approach.\\ 
Connections to free quantum field theory are
discussed for both motivation and physical
interpretation. It is conjectured that the
necessary spectral data can be generically
obtained from an effective field theory having
the fundamental structures of generalized quantum
mechanics: a decoherence functional and a choice
of histories.
\end{abstract}

\maketitle


\section*{Introduction}

Two  experimentally  verified  theories 
describe  at present the physical world: quantum
field theory and general relativity. Both have 
been  extremely  successful   in  their  tested 
ranges  of applicability.

Quantum field  theory, particularly implemented
in  the so-called standard model,  describes the
types  and behavior of  elementary particles as
measured in accelerator  experiments and as
experienced by everyday contact with matter.

General relativity  is concerned with the 
classical spacetime in which quantum  field
theory takes  place. This spacetime,  a
four-dimensional Lorentzian manifold with events
being its points, can have a complicated
structure both locally and globally and can be
influenced by the presence of classically
understood matter.

Both quantum  field theory as  applied in the 
standard model and general  relativity indicate 
intrinsically that  they cannot  be valid under
very extreme circumstances. But moreover they are
not fully  compatible even  under  rather  usual
conditions  with the problem  being  that 
matter  is  described  by  a quantum theory
whereas  spacetime   interacting  with  the  
quantum  matter  is described classically.

For all these  reasons it is believed that  it
should be possible to find a more advanced theory
in which also gravity is quantized and  in  
which  both  the  compatibility   problem  for 
general relativity and  quantum field theory and 
their internal problems are resolved.

Such  theories have  already been  proposed,
most  notably string theory \cite{Witten96}.
While the internal consistency of such a theory
turns out to be a  difficult problem, another
issue arises  once the theory is formulated. How 
can one relate it to  the physical world? The
interpretational side of a physical  theory has
at certain points of history not  been trivial
but here the  question stands with a new
urgency.  Practically all measurements that  are
performed in experimental  physics  use 
implicitly  the  notion of classical spacetime. 
The  measurements  of  positions  and  times 
play  a dominant role,  and there is  a clear
practical  understanding of them.  But in  a
theory  where gravity  is quantized,  there is no
classical spacetime in its  postulates. The
obvious conclusion is that unless one  is able to
recover from  such a theory classical spacetime,
at least in an approximative  sense, the theory
may be a nice  piece of  mathematics but  does
not  make contact  with the physical world and is
as a physical theory rather useless.

This  work  is  concerned  with  providing  a
tool for recovering classical spacetime from an
advanced  theory and is thus aimed at the
interpretation of a quantum  theory of gravity.
It is assumed here that such  a theory can first
be  simplified to an effective low  energy
theory  which will  look like  a usual quantum
field theory  but without  having specified 
spacetime yet.  In such  a situation no 
Lorentzian manifold is  present, but there  are
many structures that  contain what one can  call
spectral information. It comes from the structure
of  the algebra of observables of the effective 
theory   and  eventually  from   structures 
like  the decoherence  functional  of 
generalized  quantum  mechanics. The problem is
thus to describe classical spacetime by spectral
data.

There  is a  theory doing   just that  for
Riemannian  spaces: A. Connes'  
noncommutative    geometry
\cite{Connes94,Connes95}.   Noncommutative  
geometry describes classical  spaces by
commutative  algebras of functions on them
together  with some additional structures on 
them. It is actually  more  powerful  than  is 
needed  here:  Noncommutative geometry is  able
to deal  even with noncommutative  algebras not
corresponding  to any  classical space.  In an 
indirect way this fact is actually useful even 
in the present situation where only a classical
space is wished for: The understanding of the
general noncommutative  case  is  more  direct 
in  separating  out which concepts  are  of 
fundamental  importance  and  which  are from a
broader   perspective   just   particularities.  
One   structure recognized in this way as being
important, the spectral triple, will be
especially useful in the considerations presented.

So  in  a  more  specific  view  the  problem 
is  to discuss how noncommutative geometry can
be  used   to  describe  spacetime in the
particular commutative case.   Unfortunately,
the  present mathematical  framework  is  able 
to  deal  only  with spaces of Riemannian type, 
having a nonnegative  distance between any  two
points.  There  it  is  very  efficient  in 
using spectral data: Practically all the
geometric information  is contained in just a few
relatively  simple structures. The  question is
whether  the same is possible in the Lorentzian
case.

The answer to this question is  the main topic
and result of this work.  Compared to  Riemannian
spectral  geometry there  is a new phenomenon 
recognized:  causal  relationships.  Inspired  by
the thorough discussion  of causality in  quantum
field theory  
\cite{Haag-Kastler,Streater-Wightman,Haag,Yurtsever1,Yurtsever2},  its place in the framework of 
noncommutative geometry is found. With this
understanding it  is possible to show that 
again, as in Riemannian  geometry,  the 
spectral  data  exhibit  a  remarkable efficiency
in the description of Lorentzian spaces, at least
if they are globally hyperbolic which will be
assumed throughout.

This  gives hope  that the  adopted approach  may
turn  out to be actually useful  in the way  it
is wished  to be useful  from the physical
context. Several remarks and conjectures on
applications in  physical  interpretations  are 
put  forward.  Many technical questions are left
open for further considerations but have now a
clearer  formulation  and  context   and  can 
thus  be  attacked gradually.

The work is organized in the following way:

Section  \ref{FieldSection} discusses a free Weyl
spinor field and the fermionic quantization of
its covariant phase space. 

The local and causal structures of this field
theory are emphasized in Section
\ref{LocalObservables}. 

Section  \ref{ConnesSpectralTriple} reviews
briefly the spectral triple of Connes' spectral
geometry.

A na\"{\i}ve  spectral  description of
Lorentzian  globally hyperbolic  manifolds is
given in Section \ref{SNG}. 

In Section \ref{LorentzSpectralData} the
information contained in causal relationships is
discussed and used to obtain a rather compact
description of  spacetime. The view obtained is
the main result of this work.

This is summarized in the conclusion.


\section{Fermionic quantization of free Weyl 
spinor fields} \label{FieldSection}

The field theory considered in this paper will be
that of a fermionic Weyl spinor field. This choice
is maybe not overly surprising in view of the
role played by such fields in the standard model
of particle physics. The primary motivation is,
however, the importance of spinor fields in
spectral geometry as will become clear in
Section  \ref{ConnesSpectralTriple}.

The covariant phase space $\mathcal{S}$ of a
classical free Weyl spinor field $\psi$ is the
linear space of solutions of the equation of
motion following from the action $S$,
\begin{align}
S[ \psi ] &= Re \int_{\Omega }{\bar{\psi}  D \psi
d\Omega},\label{Action}\\
\intertext{i.e., the Dirac equation}
 D\psi &= 0.\label{DiracEquation}
\end{align}
Here $ D$ is the Dirac operator, $\bar{\psi }$ is
the Dirac adjoint of $\psi $
\cite{Budinich-Trautman} and $\Omega$ is an
arbitrarily chosen region of spacetime. 

If the spacetime manifold $M$ is assumed to be
globally hyperbolic (i.e.,  is topologically
$\Sigma \times \mathbb{R} $,  sliced by spacelike
Cauchy surfaces diffeomorphic to the
3-dimensional manifold $\Sigma $ \cite{Wald})
then there is a Hermitean inner product $s$ on
the space of solutions $\mathcal{S}$ of the Dirac
equation expressed as an integral over a
spacelike Cauchy surface $\Sigma $,
\begin{align}
\bar{\phi }\circ s \circ \psi = \int_{\Sigma
}{\bar{\phi }{\gamma }_{\mu}\psi
{d\Sigma }^{\mu }}\label{DiracProduct}
\end{align} Here ${d\Sigma }^{\mu }$ is the
future directed hypersurface element induced from
the spacetime volume element $d\Omega $. In order
for $s$ to be a Hermitean inner product on the
space of solutions $\mathcal{S}$, it has to be
independent of the choice of $\Sigma $. Indeed,
given two spacelike Cauchy hypersurfaces
${\Sigma}_{1}$ and ${\Sigma}_{2}$, the difference
in the corresponding Hermitean inner products can
be by Stokes' theorem expressed by a spacetime
integral over the region $\Omega $ enclosed by
${\Sigma}_{1}$ and ${\Sigma}_{2}$, vanishing in
consequence of the equation of motion
\ref{DiracEquation}:
\begin{align}
\int_{{\Sigma }_{1} }{\bar{\phi }{\gamma
}_{\mu}\psi {d\Sigma }^{\mu }}-\int_{{\Sigma
}_{1} }{\bar{\phi }{\gamma }_{\mu}\psi {d\Sigma
}^{\mu }}= \int_{\Omega }{\left( \bar{\phi} D
\psi - \overline{ D \phi}\psi\right) d\Omega }
\end{align}
The real part $\mu $ of the Hermitean inner
product $s$,
\begin{align}
\bar{\phi }\circ \mu \circ \psi = Re \int_{\Sigma }
{\bar{\phi }{\gamma }_{\mu}\psi
{d\Sigma }^{\mu }} 
\end{align}
is a real bilinear symmetric inner product on the phase 
space  $\mathcal{S}$. Its inverse is the fermionic causal 
Green's function ${\tilde{G}}_{F}$.

It can be shown \cite{DeWitt65} that the
fermionic causal Green's function
${\tilde{G}}_{F}$ has the meaning of the Poisson
bracket $\{\bullet , \bullet\} $ of classical
mechanics \cite{DeWitt83}:
\begin{align}
\{\bullet , \bullet\} = {\tilde{G}}_{F} =
{\mu}^{-1}
\end{align}

Once the classical description of a system (e.g.
a field) is known, one can make an educated guess
of what the correct quantum theory is, i.e., one
can quantize the field theory.  In principle
there are two rather different ways to do that,
namely quantization by path integrals
\cite{Feynman-Hibbs} and  canonical quantization
(see, e.g.\cite{DeWitt65,DeWitt83,Woodhouse}).
Here the latter is chosen, since it leads more
directly to an algebraic setting used in
noncommutative geometry.

In fermionic canonical quantization, chosen in
agreement with the spin-statistics
theorem\cite{Streater-Wightman}, one starts with
the classical phase space $\mathcal{S}$ equipped
with the symmetric inner product  $\mu $. The
functions on the classical phase space
$\mathcal{S}$, the classical observables, are
then replaced by elements in a noncommutative
algebra, the algebra of observables following
some rules which turned out to be useful in
particular cases. The rules are as follows:

First, a special set $F(\mathcal{S})$ of function
on the phase space has to be selected. The set
$F(\mathcal{S})$ of chosen classical observables
should be closed under taking the Poisson bracket
$\{\bullet,\bullet\}$, i.e.,
\begin{align}
\{ a, b\} \in F(\mathcal{S}) &&\text{ for $a,
b\in F(\mathcal{S})$}, 
\end{align}
	Second, a linear map $\hat{\psi}$ into a
	complex associative algebra $\mathbf{A}$
	should be given,
\begin{align}
\hat{\psi}:F(\mathcal{S}) &\rightarrow
\mathbf{A}.
\end{align}
The map $\hat{\psi}$ should satisfy a commutation
relation replacing the Poisson bracket by a
commutator:
\begin{align}
\hat{\psi}(a)\hat{\psi}(b) + \hat{\psi}(b)\hat{\psi}(a) =
 i\hat{\psi}(\{ a, b \} )              \text{ for all $a, b\in F(\mathcal{S})$},\label{AnticommutationRelation}
\end{align}
and its image $\hat{\psi}(F(\mathcal{S}))$ should
generate the algebra $\mathbf{A}$.
\begin{note}
If $F(\mathcal{S})$ contains the constant
functions on $\mathcal{S}$ (which have vanishing
Poisson brackets with all other functions on
$\mathcal{S}$), then their image under the
mapping $\hat{\psi}$ must be in the centre of the
algebra $\mathbf{A}$, and if $\mathbf{A}$ is
central then the image of constant functions is
proportional to the unit $\mathbf{1}$ in the
algebra. A not very surprising addition to the
quantization rules then usually is the
requirement
\begin{align}
\hat{\psi}(k) = k\mathbf{1} &&\text{for all constant 
functions $k$ on $\mathcal{S}$}.
\end{align}
\end{note}
In general, one of the difficulties of these
rules is the potentially complicated
anticommutation relation
(\ref{AnticommutationRelation}), and another is
the choice of $F(\mathcal{S})$. Obvious choices,
like the space of all continuous functions on
$\mathcal{S}$, are plagued by inconsistencies or
by giving an algebra that is far too big compared
with the one that gives a quantum theory in
agreement experiment. To deal with this
situation, additional information is usually
necessary (see e.g. \cite{Woodhouse}), and even
then it is a difficult problem.  The situation
radically simplifies for a free system (i.e. one
with a linear phase space $\mathcal{S}$) as the
one considered here. The correct choice of
$F(\mathcal{S})$ is then the space of linear
observables.

One can define the field operator $\Psi (f)$ for
a classical solution $f \in \mathcal{S}$
\begin{align}
\Psi (f) =\hat{\psi}(\mu \circ
f),\label{SolutionField}
\end{align}
and write the anticommutation relation
(\ref{AnticommutationRelation}) in the form 
\begin{align}\label{SolutionAnticommutationRelation}
\Psi (f)\Psi (g)-\Psi (g)\Psi (f)=i
(f\circ\mu\circ g) \mathbf{1} &&\text{for $f,g\in
\mathcal{S}$.}
\end{align}

The ${C}^{\ast}$-algebra of observables of the
quantum field generated from this anticommutation
relation is unique and independent of a
completion of $\mathcal{S}$
\cite{Plymen-Robinson,Bratteli-Robinson2}. It has
a unique minimal enveloping von Neumann algebra
\cite{Plymen-Robinson} having, up to unitary
isomorphism, a unique regular irreducible
representation by bounded operators in a Hilbert
space. There is no information whatsoever in this
algebra about the smooth structure of spacetime.


\section{Local algebras of
observables}\label{LocalObservables}

If the ${C}^{\ast}$-algebra of observables is
considered by itself, without reference to its
origin, then it is sufficient to express the
evolution of the field by automorphisms and the
space of states by normed positive linear
functionals (see \cite{Bratteli-Robinson1}), but
then the physical interpretation is completely
lost. 

A somewhat similar loss of interpretation can be
observed if a classical system is judged on the
basis of its phase space only, where canonical
transformations can rather arbitrarily change the
meaning of coordinates and momenta. It is
possible to argue that, e.g., the topology of the
phase space is specific to the system, but this
is by no means sufficient to give a complete
description if there actually is a fundamental
distinction between coordinates and momenta. 

As mentioned in Section \ref{FieldSection}, the
algebra of observables does in this case not
contain any information about spacetime. 

Some structure has thus to be given to the
algebra of observables of a quantum field in
order to enable one to give its physical
interpretation. One could, of course, just
remember the whole construction of the algebra of
observables, starting with the classical field.
In a path integral approach this would not be so
bad, since classical histories are part of that
framework, but in an algebraic approach to
quantum field theory, where the classical field
has just the position of an effective
approximation, this is definitely not what one
would wish to do.  The widely accepted solution
is to give the algebra of observables the
structure of a local algebra
\cite{Haag,Bratteli-Robinson1}. The idea is to
associate with each region of spacetime $\Omega$
a subalgebra $\mathbf{A}(\Omega)$ of the algebra
$\mathbf{A}$ of observables. Thus one obtains a
set of subalgebras indexed (not necessarily
unambiguously) by the set $I$ of open subsets of
spacetime. 

For many technical purposes it is not necessary
to keep the reference to spacetime, and only some
properties of the index set $I$ are extracted and
required. This is the case of the definition of a
quasi-local algebra
\cite{Haag,Bratteli-Robinson1}. However, since
here interpretation is the main concern,  the
full link to spacetime will be required
\cite{Yurtsever1,Yurtsever2}.

\begin{definition}
A ${C}^{\ast}$-algebra $\mathbf{A}$ together with 
a spacetime manifold $M$ is local if the following
three conditions all hold:
\begin{enumerate}
\item{For each open subset $\Omega$ of $M$ there
is a central ${C}^{\ast}$-algebra
$\mathbf{A}(\Omega)$, with
$\mathbf{A}(\emptyset)=\mathbb{C}$, and
$\mathbf{A}(M)= \mathbf{A}$.}
\item{For any collection $\{ {\Omega}_{i}\}$ of
open subsets of $M$ one has 
\begin{align}
\mathbf{A}\left( \cup_{i} {\Omega}_{i}\right) =
\overline{\langle\cup_{i} \mathbf{A}\left(
{\Omega}_{i}\right)\rangle}\notag
\end{align}
(On the right hand side of this equation is the
closure of the algebraic envelope
$\langle\cup_{i} \mathbf{A}\left(
{\Omega}_{i}\right)\rangle $ of $\cup_{i}
\mathbf{A}\left( {\Omega}_{i}\right) $.)
}
\item{If the regions ${\Omega}_{1}$,
${\Omega}_{2}$ are not in causal contact, then
the corresponding algebras $\mathbf{A}\left(
{\Omega}_{1}\right)$, $\mathbf{A}\left(
{\Omega}_{2}\right)$ commute in the Bose case and
graded-commute in the Fermi case.} 
\end{enumerate}
\end{definition}

\begin{example}
The quantized Weyl spinor field can be given the
structure of a local algebra. The Green's
function ${\tilde{G}}_{F}$ of the field can be
used to produce from any smooth density $\nu $ on
the spacetime manifold $M$ a solution $f$:
\begin{align}
{f}^{p} = {({\tilde{G}}_{F})}^{pq}
{\nu}_{q}\label{MeasureFieldFermi}
\end{align}
and to each solution $f$ one can by
(\ref{SolutionField}) associate a quantum
observable $\Psi (f)$. Given a subset $\Omega$ of
spacetime, the algebra $\mathbf{A}(\Omega)$ can
be then generated by densities with support in
$\Omega$. If the supports of two measures ${\nu
}_{1}$, ${\nu }_{2}$ are not causally connected,
then the corresponding classical solutions
${f}_{1}$, ${f}_{2}$ can be checked to have a
vanishing product  ${f}_{1}\circ\mu\circ
{f}_{2}$, and the corresponding quantum
observables $\Psi ({f}_{1})$, $\Psi ({f}_{2})$
thus anticommute. 
\end{example}

	A pleasant feature of the local algebra
	structure is that the
	${C}^{\ast}$-subalgebras
	$\mathbf{A}(\Omega)$ (with $\Omega\in M$)
	of $\mathbf{A}$ are actually sufficient
	to reconstruct the spacetime $M$ as a
	topological space and to determine its
	causal structure, as observed by
	U.Yurtsever \cite{Yurtsever1,Yurtsever2}.
	
\section{Connes' spectral triple}\label{ConnesSpectralTriple}

A geometric space may be described by its set of
points with some additional structures, or,
alternatively, by the algebra of functions on it,
again with some additional structures. The first
point of view is the one of classical geometry.
The second may be taken as a starting point for a
far more general and powerful theory, A. Connes'
noncommutative geometry \cite{Connes94}, and is
adopted here. In particular, a space can be
encoded in the form of a spectral triple
\cite{Connes95}.

\begin{definition}\label{SpectralTriple}
A {\em spectral triple} $(\mathbf{A},\mathcal{H},D)$ is given 
by an involutive algebra of operators $\mathbf{A}$ in a Hilbert 
space $\mathcal{H}$ and a selfadjoint operator 
$D={D}_{\ast}$ in $\mathcal{H}$ such that
\begin{enumerate}
\item{The resolvent ${(D-\lambda)}^{-1}, \>
\lambda\not\in\mathbb{R}$, of $D$ is compact}
\item{The commutators $[D,a]=Da-aD$ are bounded,
for any $a\in\mathbf{A}$}
\end{enumerate}
The triple is said to be {\em even} if there is a
hermitean grading operator $\gamma $ on the
Hilbert space $\mathcal{H}$ (i.e.
${\gamma}^{\ast}=\gamma,\>
{\gamma}^{2}=\mathbf{1})$ such that 
\begin{align}
\gamma a &= a \gamma &\text{ for all $a\in \mathbf{A}$}\\
\gamma D &= - D \gamma &
\end{align}
Otherwise the triple is called {\em odd}
\end{definition}

\begin{note}
This section is only concerned with introducing
the spectral triple and mentioning its properties
to be used in the applications. From that it is
not fully clear why one should be interested in
exactly this kind of structure,  so some
motivation is clearly missing here. See however
\cite{Connes94,Connes95} for the deep and solid
structure of noncommutative geometry that is
supporting the spectral triple.   
\end{note}

The following example is of great importance.

\begin{example}\label{DiracSpectralTriple}
On a compact Riemannian spin manifold $M$ there
is canonically the following spectral triple
$({C}^{\infty }(M) ,{L}^{2}(M,S),D)$, the Dirac
triple \cite{Connes94}, \cite{Connes95}. Here
${C}^{\infty }(M) $ is the commutative algebra of
smooth complex functions on $M$, ${L}^{2}(M,S)$
is the Hilbert space of square integrable
sections of the complex spinor bundle $S$ over
$M$ and D is the Dirac operator. The algebra of
functions ${C}^{\infty }(M) $  acts on the
Hilbert space ${L}^{2}(M,S)$ by pointwise
multiplication
\begin{align}
(f\psi ) (p) &= f(p)\psi (p) &\text{for all 
$f\in {C}^{\infty }(M)
,\psi \in {L}^{2}(M,S), p\in M$}\\
\intertext{and the commutator with the Dirac
operator $D$ with a function $f$ is}
[D,f]&=\gamma df &\text{for $f\in {C}^{\infty
}(M) $.} 
\end{align}
$\gamma $ is the Clifford map from the cotangent
bundle into operators on ${L}^{2}(M,S)$. 
\end{example}

In Example \ref{DiracSpectralTriple} the algebra
was taken to be ${C}^{\infty }(M)$. Such a choice
contains a lot of information and is actually not
necessary. In the definition of the Dirac
spectral triple it is sufficient to take instead
of ${C}^{\infty }(M)$ any algebra $\mathbf{A}$
that has the same weak closure (double commutant)
${\mathbf{A}}^{''}$ as has ${C}^{\infty }(M)$.
Such an algebra does not necessarily contain any
information about the topology or differential
structure of $M$ whatsoever. From $\mathbf{A}$
alone only $M$ as a set of points can be obtained
as the spectrum of $\mathbf{A}$. The rest,
however, can then be recovered from the structure
of the spectral triple including the notion of
smooth functions and Lipschitz functions.
Lipschitz functions with  Lipschitz constant $1$
can then be used to define a distance function
$d$ on $M$. This means that a Riemannian spin
manifold can be replaced by a spectral triple
without the loss of any information about it. The
facts are summarized in Proposition
\ref{DiracSpectralTripleProposition} (see
\cite{Connes95}).  

\begin{proposition}\label{DiracSpectralTripleProposition}
Let $(\mathbf{A},{L}^{2}(M,S),D)$ be the Dirac
spectral triple associated to a compact
Riemannian spin manifold M. Then the compact
space $M$ is the spectrum of the commutative
${C}^{\ast}$-algebra norm closure of
\begin{align}
{\mathbf{A}}_{B}&= \{ a \in {\mathbf{A}}^{''}\mid [D,a] 
\text{ bounded}\} &\\
\intertext{while the geodesic distance $d$ on $M$
is given by}
d(p,q) &=\sup{\{ \mid f(p)-f(q)\mid ;
 f\in {\mathbf{A}}_{B}, \parallel [D,f]\parallel\leq 1 \} 
}\label{DistanceFunction} 
\end{align} 
\end{proposition}

It is now in question whether one can reconstruct
 from a spectral triple a manifold if one is not
assured that the spectral triple actually comes
 from a manifold. With some additional conditions
it will certainly be possible to prove in the
future a theorem in this direction. One helpful
tool for this purpose is a real structure $J$ on
the spectral triple \cite{Connes95},
\cite{Connes96a}. 

\begin{example}
In the case of the Dirac spectral triple of
Example \ref{DiracSpectralTriple} a real
structure is given by the charge conjugation
composed with complex conjugation (see
\cite{Budinich-Trautman}).
\end{example} 

Before giving its general definition it should be
mentioned that for simply connected spaces the
real structure ensures that the spectrum of a
spectral triple will have the homotopy type of a
closed manifold \cite{Connes95},
\cite{Connes96a}. In addition to that, its
dimension is governed by the spectrum of the
Dirac operator \cite{Gilkey}. So a theorem
examining which commutative spectral triples are
classical Riemannian manifolds is not out of
sight. The considerations of the next sections
would be best motivated by such a theorem but
making use of it is at this
point probably premature.

\begin{definition}
A real structure $J$ on the spectral triple
$(\mathbf{A},\mathcal{H})$ is an antilinear
isometry $J$
\begin{align}
J:\mathcal{H}&\rightarrow \mathcal{H}&\\
\intertext{ such that} 
Ja{J}^{-1} &= {a}^{\ast} &\text{for all
$a\in\mathbf{A}$}\\
{J}^{2} &= \epsilon &\\
JD &= {\epsilon}^{'}DJ &\\
J\Gamma &= {\epsilon }^{''}\Gamma J
\end{align}
where the signs $\epsilon, {\epsilon}^{'}, 
{\epsilon}^{''}\in \{ -1, +1\} $ are given by the following 
table with $\nu$ being the dimension of the space $mod\> 8$:
\begin{equation}\label{RealStructureTable}
\begin{array}{|c|c|c|c|c|c|c|c|c|}
\hline
\nu & 0 & 1 & 2 & 3 & 4 & 5 & 6 & 7 \\ 
\hline
\epsilon & 1 & 1 &-1 &-1 &-1 &-1 & 1 & 1\\
\hline
{\epsilon}^{'} & 1 &-1 & 1 & 1 & 1 &-1 & 1 & 1\\
\hline
{\epsilon}^{''} & 1 &  &-1 &  & 1 &  &-1 & \\
\hline
\end{array}
\end{equation}
\end{definition} 

\begin{note}
The sign ${\epsilon}^{''}$ in Table
(\ref{RealStructureTable}) is shown for even
dimensions only, since for Riemannian spin
manifolds only in that case the grading
(helicity) operator ${\Gamma}$ preserves the
irreducible spin representation and has thus a
good meaning in it. In the odd case it is assumed
that only one of the two irreducible
representations is chosen and since $\Gamma$ 
switches between the two irreducible
representations it has no meaning just in one of
them. More details on spinors can be found in
\cite{Budinich-Trautman}. Also the periodicity
$mod\> 8$ of Table (\ref{RealStructureTable}), a
manifestations of the spinorial chessboard is
explained there.
\end{note}


\section{Spacetime in spectral
geometry}\label{SNG}

\begin{figure}\label{Slicing}
\epsfxsize= 5 in
\epsfbox{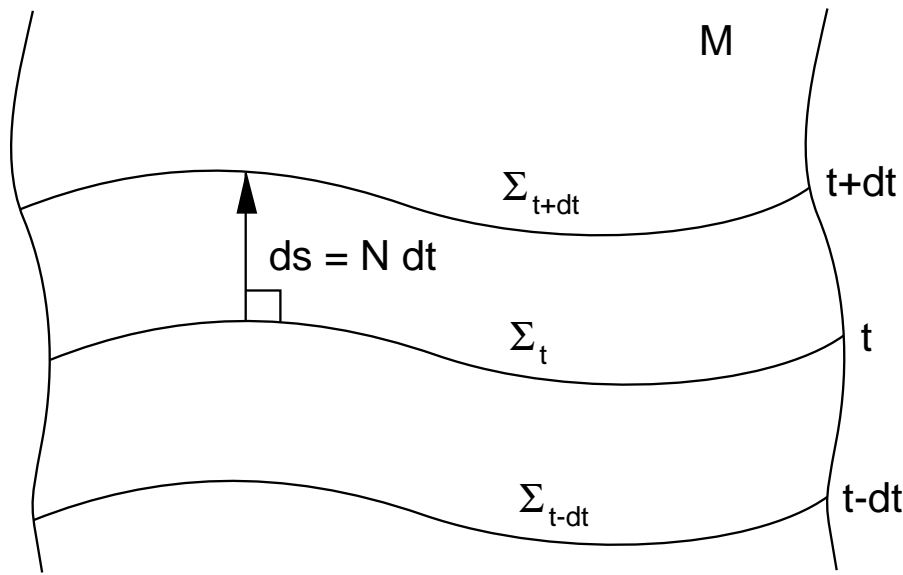}
\caption{A Cauchy foliation. The globally
hyperbolic manifold $M$ can be sliced by
spacelike Cauchy surfaces ${\Sigma}_{t}$. Each of
them can be characterized by a Dirac spectral
triple $\left( {L}^{\infty}({\Sigma}_{t}),
{L}^{2}({\Sigma}_{t},S), {D}_{t} \right)$ with
${L}^{\infty}({\Sigma}_{t})$ being the algebra of
essentially bounded functions on ${\Sigma}_{t}$,
${L}^{2}({\Sigma}_{t},S)$ being the spinor bundle
over ${\Sigma}_{t}$ and ${D}_{t}$ being the Dirac
operator on ${\Sigma}_{t}$. The normal distance
between infinitesimally close Cauchy surfaces
${\Sigma}_{t}$, ${\Sigma}_{t+dt}$ is
characterized by the lapse function $N$ on
${\Sigma}_{t}$. $N$ can be thought of as an
element in the algebra 
${L}^{\infty}({\Sigma}_{t}) = {\left(
{C}^{\infty}({\Sigma}_{t}) \right)}^{''}$, the
double commutant of the algebra of smooth
functions.}
\end{figure}

Here a Lorentzian globally hyperbolic spacetime
manifold will be characterized by spectral data.
This cannot be done directly by Connes' spectral
triple (see Definition \ref{SpectralTriple})
since it is well suited for the description of 
generalized Riemannian spaces only. This is
obvious, e.g., from the distance function
(\ref{DistanceFunction}), which cannot be
negative. A simple idea to avoid this difficulty
is to foliate the spacetime $M$ by a family of
spacelike Cauchy slices ${\Sigma}_{t}$ with
$t\in\mathbb{R}$ a coordinate time (see Figure
\ref{Slicing}). Each hypersurface ${\Sigma}_{t}$
is then Riemannian and can be characterized by a
family of Dirac spectral triples $\left(
{L}^{\infty}({\Sigma}_{t}),
{L}^{2}({\Sigma}_{t},S), {D}_{t} \right)$ (see
Example \ref{DiracSpectralTriple} and Proposition
\ref{DiracSpectralTripleProposition}) together
with some additional information on how the
spacelike slices ${\Sigma}_{t}$ are related to
each other. In particular, the normal distance
between two infinitesimally close Cauchy surfaces
${\Sigma}_{t}$ is encoded by the lapse function
$N$ (see \cite{MTW} and Figure \ref{Slicing}).
The only further information needed is the
identification ${i}_{t}:{\Sigma}_{t}\rightarrow
{\Sigma}_{0}$ of points which lie on the same
curve normal to the hypersurfaces. This can be
established in the spectral data by specifying an
automorphism
${i}_{t}^{\ast}:{L}^{\infty}({\Sigma}_{0})\rightarrow
{L}^{\infty}({\Sigma}_{t})$ 

Since the square integrable sections of the spin
bundles over the Cauchy surfaces ${\Sigma}_{t}$,
$t\in\mathbb{R}$ are valid Cauchy data for weak
solutions of the equation of motion of a Weyl
spinor field on M, there is a preferred
isomorphism between the spin bundles
${L}^{2}({\Sigma}_{t},S)$ and the space of
solutions $\mathcal{S}$ of Weyl spinors. This
means that all spectral triples can be understood
to share the same Hilbert space $\mathcal{S}$.

Summarizing, a globally hyperbolic spacetime can
be described using spectral data by
\begin{itemize}
\item{a family of spectral triples
$({\mathbf{C}}_{t}, \mathcal{S}, {D}_{t})$ with
${\mathbf{C}}_{t}$ a commutative algebra of
bounded operators on $\mathcal{S}$ and ${D}_{t}$
Hermitean (possibly unbounded) on $\mathcal{S}$}
\item{a family of lapse functions
${N}_{t}\in{\mathbf{C}}_{t}$ }
\item{an automorphism ${i}^{\ast}$ between any
two of the commutative algebras
${\mathbf{C}}_{t}$}
\end{itemize}

\begin{note}
Usually it is not required that the
identification of Cauchy surfaces has to be done
along normal lines. Then the deviation of of the
direction of identification from the normal one
has to be characterized by a shift vector field
$\vec{N}$ on the Cauchy surfaces \cite{MTW}. The
restriction to the case $\vec{N}=0$ here avoids
the necessity of a replacement of vector fields
by spectral concepts.
\end{note}

The above description agrees with  \cite{Hawkins}
except that there the automorphism ${i}^{\ast}$
is omitted. That omission seems to make the
spectral data appear incomplete from the point of
view presented there.

It is now possible to describe the quantum field
theory for Weyl spinors on the spacetime
specified by the spectral data. Since the Hilbert
space $\mathcal{S}$ in the spectral data is taken
to be the space of classical solutions equipped
with the canonical Hermitean inner product
(\ref{DiracProduct}),  this is entirely trivial:
The quantum field algebra of observables is just
the Clifford algebra generated from $\mathcal{S}$
by the anticommutation relation
(\ref{SolutionAnticommutationRelation}).

This completes the discussion of quantum field
theory on spacetime using a spectral approach but
not taking in account the causal structure
information present in the problem. This is a
natural place to reflect on the above with a few
comments.

{}From the point of view of the motivations, one
would wish to start from an algebra of quantum
observables, to specify the spectral data, and
then to construct, if possible, classical
spacetime. Such an approach will however bring
rather difficult problems: At least in the cases
where one hopes to obtain a spacetime that is a
topological or smooth manifold, one would wish to
have the one-parameter family ${\mathbf{C}}_{t}$
in some sense continuous or smooth. (It may be
viewed as a  continuous or smooth algebra bundle
over $\mathbb{R}$). This is an important, but on
the other hand technical, issue. Instead of
discussing it satisfactorily, the treatment will
rely on the case studied here starting with a
classical spacetime, producing the space of
solutions $\mathcal{S}$ of the Weyl spinor field
on it and obtaining by quantization the field
algebra $\mathbf{A}$. Then all the facts can be
viewed backwards, starting with the field algebra
$\mathbf{A}$. This is clearly dishonest to the
motivations in using as its input what should be
abandoned in the first place: classical
spacetime. On the other hand this allows one to
go through all the way from the quantum algebra
to spacetime avoiding some, in general difficult,
arguments bridged by the particular features of
this not-so-elegant example. The result is then
an understanding of what is important, and with
this, one can then gradually face the technically
difficult points. This approach has worked so far
extremely well in noncommutative geometry. In
this context, the aim here is to gain an
understanding only, thus considering the example
as a valid approach.

For a view starting from the quantum field
algebra according to the above motivations, it
would also be desirable to have a deeper
justification of the introduced structures,
particularly for the family of operator algebras
${\mathbf{C}}_{t}$ and the family of operators
${D}_{t}$ on the space $\mathcal{S}$ generating
the algebra of observables $\mathbf{A}$. It will
be suggested here in the form of two conjectures
that this may eventually be possible.

\begin{conjecture}\label{1stConjecture}
Another way to look at the family of commutative
algebras ${\mathbf{C}}_{t}$ will be offered now.
For a given value of the parameter $t={t}_{0}$,
the algebra ${\mathbf{C}}_{{t}_{0}}$ splits the
space $\mathcal{S}$ into orthogonal subspaces by
spectral projections. On the quantum level this
means that the field algebra $\mathbf{A}$ is
given preferred mutually commuting subspaces. In
the case in which the Hilbert space is finite
dimensional, these spaces are complex one
dimensional. It is conjectured that this
structure is sufficient to determine a preferred
complete set of commuting projectors in the
algebra of observables $\mathbf{A}$ or eventually
in its (unique) minimal enveloping von Neumann
algebra. If that is the case, then the choice of
${\mathbf{C}}_{{t}_{0}}$ may be understood as the
choice of a set of histories in generalized
quantum mechanics
\cite{Hartle,Isham,Isham-Linden,Isham-Linden-Schreckenberg}.
This would to a large degree justify the
introduced structures from a very fundamental
point of view.
\end{conjecture}

\begin{conjecture}\label{2ndConjecture}
If Conjecture \ref{1stConjecture} is in some way
correct, then the family ${D}_{t}$ of Hermitean
operators on $\mathcal{S}$ can be recovered from
the decoherence functional of generalized quantum
mechanics on histories of the quantum field
$\mathbf{A}$. 
\end{conjecture}

These conjectures are a topic of future research.
They are stated here only to show that what was
reached so far is really following the call of
the motivations put forward in the Introduction,
which would not be so easy to see otherwise.


\section{Spectral data and the causal structure
of spacetime.} \label{LorentzSpectralData}

The spectral data describing spacetime as
presented in the previous section are sufficient.
But they do not take into account the fact that
causal structure information is also stored in
the family of spectral triples in a way that was
not yet exploited.

\begin{figure}\label{CausalityFigure}
\epsfxsize= 5 in
\epsfbox{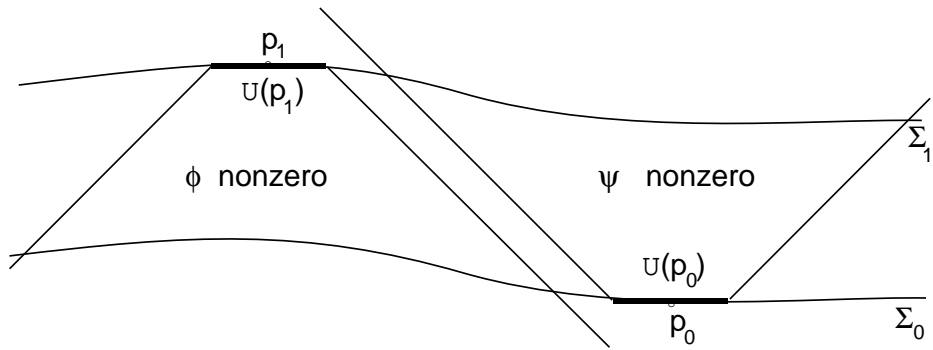}
\caption{Causal contact. Any solution $\psi$ with
Cauchy data on ${\Sigma}_{0}$ supported in
$\mathcal{U}({p}_{0})$ has a vanishing inner
product with any  solution $\phi$ with Cauchy
data on ${\Sigma}_{1}$ supported in
$\mathcal{U}({p}_{1})$. The points ${p}_{0}$,
${p}_{1}$ are not causally connected.}
\end{figure}

To understand that, consider two spacelike Cauchy
surfaces ${\Sigma}_{0}$, ${\Sigma}_{1}$ on the
spacetime manifold (see Fig\-ure
\ref{CausalityFigure}). They are de\-scri\-bed by
the spec\-tral triples $({\mathbf{C}}_{0},
\mathcal{S}, {D}_{0})$, $({\mathbf{C}}_{1},
\mathcal{S}, {D}_{1})$. Given two points
${p}_{0}$, ${p}_{1}$ on these Cauchy surfaces
(${p}_{0}\in {\Sigma}_{0}$, ${p}_{1}\in
{\Sigma}_{1}$) it is now possible just to decide
whether they are in causal contact or not. If and
only if the points ${p}_{0}$, ${p}_{1}$ are not
in causal contact, the value of the Weyl spinor
field at the point ${p}_{0}$ cannot influence the
value of the field at the point ${p}_{1}$. In
more precise terms  on can say that there exist
open neighborhoods $\mathcal{U}({p}_{0})$,
$\mathcal{U}({p}_{1})$ of the points ${p}_{0}$,
${p}_{1}$ in ${\Sigma}_{0}$, ${\Sigma}_{1}$ such
that any solution $\psi$ of the equation of
motion of the Weyl spinor field with Cauchy data
on ${\Sigma}_{0}$ supported in
$\mathcal{U}({p}_{0})$ has a vanishing inner
product with any  solution $\phi$ with Cauchy
data on ${\Sigma}_{1}$ supported in
$\mathcal{U}({p}_{1})$. To identify solutions in
$\mathcal{S}$ which have Cauchy data on
${\Sigma}_{i}$ supported in a certain region
$\mathcal{U}({p}_{i})\subset {\Sigma}_{i} $ from
the spectral data is easy: they are just given as
elements of the ranges of the spectral projection
corresponding to $\mathcal{U}({p}_{i})$.

\begin{note}
If one is willing to use generalized eigenvectors
then causal contact can be expressed in the
following way. A (generalized) solution with
Cauchy data on ${\Sigma}_{0}$ supported in the
point ${p}_{0}$ is a generalized eigenvector of
the algebra ${\mathbf{C}}_{0}$ satisfying
\begin{align}
a\psi = a({p}_{0}) \psi &&\text{for
$a\in{\mathbf{C}}_{0}$,}
\end{align}
with $a({p}_{0})$ being the value of the function
$a$ at the point ${p}_{0}$. The vector $\psi$ can
then be for briefness called an eigenvector of
point ${p}_{0}$. Then two points are not in
causal contact if and only if all their
eigenvectors are orthogonal. 
\end{note}

One can now summarize:

\begin{observation}\label{Observation}
Using the family  ${\mathbf{C}}_{t}$ of
commutative algebras represented on the Hilbert
space $\mathcal{S}$ of solutions, one can recover
spacetime as a set of points and find by the
above procedure which points are in causal
contact, using the Hermitean inner product on
$\mathcal{S}$. 
\end{observation}

This observation is of central importance. Before
using it to reduce the spectral data necessary to
describe a Lorentzian spacetime, a two
connections will be made.

First, from the point of view of differential
equations it is not surprising that the Hermitean
inner product on $\mathcal{S}$ contains
information on the causal structure, since as
mentioned in Section \ref{FieldSection} the real
part of it is the inverse of the causal Green's
function.

Second, from the point of view of quantum field
theory the orthogonality of classical solutions
with Cauchy data locally supported around two
points  ${p}_{0}$, ${p}_{1}$ has as its
consequence (or, if one wishes, as its origin)
the graded commutativity of the corresponding
${C}^{\ast}$-subalgebras of the local algebra
$\mathbf{A}$  of observables generated from
$\mathcal{S}$. This is the point where the notion
of causality makes contact with Section
\ref{LocalObservables} and with some of the
motivations for this work given in the
Introduction.

\begin{figure}\label{LightCone}
\epsfxsize= 5 in
\epsfbox{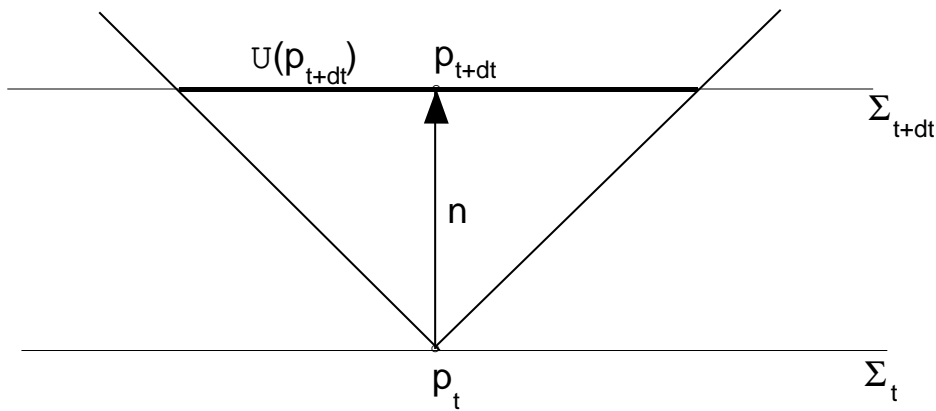}
\caption{The geometry of Cauchy surfaces, causal
contact and the geometry of spacetime. The point
${p}_{t}$ on ${\Sigma}_{t}$ has as its region of
causal contact on ${\Sigma}_{t+dt}$ the disk
$\mathcal{U}({p}_{t+dt})$ (including its bounding
sphere). The square of the radius of the sphere
is the negative of the square of the normal
spacetime distance between the Cauchy surfaces
${\Sigma}_{t}$, ${\Sigma}_{t+dt}$, and the center
${p}_{t+dt}$ of the sphere
$\mathcal{U}({p}_{t+dt})$ is the point reached by
the normal vector $n$ based in ${p}_{t}$.  }
\end{figure}

Now the consequences of Observation
\ref{Observation} will be discussed. First of
all, the family of spectral triples
$({\mathbf{C}}_{t}, \mathcal{S}, {D}_{t})$ of
Section \ref{ConnesSpectralTriple} contains
already all necessary information about spacetime
and no automorphism ${i}^{\ast}$ between the
algebras ${\mathbf{C}}_{t}$ and no lapse function
$N$ need to be specified. Indeed, by knowing the
geometry of the Cauchy surfaces ${\Sigma}_{t}$
corresponding to the spectral triples
$({\mathbf{C}}_{t}, \mathcal{S}, {D}_{t})$ and
the causal structure one can find the normal
identifications of points and the normal
distances between infinitesimally close Cauchy
surfaces (see Figure \ref{LightCone}).

Thus a large part of the spectral data can be
just left out, and the remaining family of
spectral triples gives now a quite efficient
description. But it is still considerably
redundant. To see this is not difficult: If the
metric information contained in the operators
${D}_{t}$ is omitted, then the conformal
structure of spacetime is still rigidly fixed.
But not all metrics are conformally related, and
thus the ${D}_{t}$ determining the metric on the
Cauchy surfaces cannot be chosen at will but have
to agree with the conformal structure. This means
that the spectral data of spacetime can be
further reduced. How this has to be done in a
useful way will be left for consideration in the
future. But even without that a conceptual result
is appearing: The spectral data describing a
Lorentzian manifold do so in a very efficient
way. This result based on Observation
\ref{Observation} is the main claim of this work.

\begin{note}
There is a way of giving less redundant spectral
data, if one is willing to lose metric
information and keep just the conformal structure
of spacetime. It is shown in \cite{Connes94} that
for building just differential geometry without
metric information, it is sufficient to take,
instead of the spectral triple with an unbounded
operator $D$, the same spectral triple but with
$D$ replaced by $F=sgn \>D$, the sign of the
operator $D$. This is actually a grading operator
on $\mathcal{S}$ since ${F}^{2}=\mathbf{1}$. Thus
the spectral triple $({\mathbf{C}}_{t},
\mathcal{S}, {F}_{t})$ with a family of grading
operators contains the topological and causal as
well as differential geometric information on
spacetime.
\end{note}

\begin{note}
One may wonder where the efficiency of the
spectral data in the presented description comes
 from. In the case of the spectral triple A.
Connes argued \cite{Connes94,Connes95} that most
of the information is not in the algebra of the
triple, giving basically just a set of points,
nor  in the chosen Hermitean operator, fully
described by its spectrum, but in the
relationship between them. This explanation can
be used here again: Most of the information in
the spectral data is not in the commutative
algebras ${\mathbf{C}}_{t}$ represented on
$\mathcal{S}$ but in the relationships between
them. Indeed, the strong causal structure is
purely a result of this.
\end{note}

\section*{Conclusion}

Motivated by the need to recover classical
spacetime from a theory of quantum gravity in
order to achieve the theory's physical
interpretation, the thesis examines the
possibility of describing classical Lorentzian
spacetime manifolds by spectral data.

Following in Section \ref{SNG} a na\"{\i}ve
Hamiltonian approach, the spectral data for a
Lorentzian manifold are specified as a family of
A. Connes' spectral triples with a common Hilbert
space and additional structures known from
Hamiltonian general relativity: a family of lapse
functions and an identification of Cauchy
surfaces implemented by isomorphisms of the
algebras in the spectral triples. This gives a
complete description of spacetime, trivially
extended to a free quantum field theory on
spacetime.

However, in Section \ref{LorentzSpectralData} it
is realized that the spectral description of
spacetime automatically contains unused
information on causal relationships. The use of
this information leads to a significant reduction
of the spectral data. The family of lapse
functions and the identification of Cauchy
surfaces can be completely left out, and still
there is considerable redundancy in the data
present. The discovery of the place of causal
relationships in spectral geometry thus leads to
a very efficient spectral description of
spacetime. This is the main result of this
thesis. 

With the result attained here, there are now two
well motivated problems of conceptual importance:
\begin{enumerate}
\item{The remaining redundancy in the spectral
data should be removed and the result put into a
useful form to be recognized as standard.}
\item{The way in which the result may fit into an
interpretation of quantum gravity should be
clarified, possibly along the lines of
Conjectures \ref{1stConjecture} and
\ref{2ndConjecture}}
\end{enumerate}
Moreover, there are also many further points of
technical nature, to be worked out. To suggest
just one of them as an example, it would be
desirable to have a usefully formulated
expression for spacetime distances.

With the insight obtained here, these questions
are now open to future investigations.

\section*{Acknowledgements}

The author would like to thank Pavel Krtou\v{s}, 
Don N. Page and Georg Peschke for a number of 
invaluable discussions. 


\end{document}